\documentclass[reprint, amsthm, amsmath,amssymb, superscriptaddress, aps]{revtex4-2}

\usepackage{amsthm,amsmath,amssymb}
\usepackage{graphicx}
\usepackage{cancel}
\usepackage{color}
\usepackage{chngcntr}
\usepackage{xcolor}
\usepackage{orcidlink}

\begin{document}

\title{Continuum Landau surface states in a non-Hermitian Weyl semimetal}
\date{\today}

\author{Shuxin Lin\orcidlink{0000-0002-9093-8531}}
\affiliation{School of Physical and Mathematical Sciences, Nanyang Technological University, Singapore 637371, Singapore}

\author{Rimi Banerjee\orcidlink{0000-0001-5840-2756}}
\affiliation{School of Physical and Mathematical Sciences, Nanyang Technological University, Singapore 637371, Singapore}

\author{Zheyu Cheng\orcidlink{0000-0001-5009-7929}}
\affiliation{School of Physical and Mathematical Sciences, Nanyang Technological University, Singapore 637371, Singapore}

\author{Kohei Kawabata}
\affiliation{Institute for Solid State Physics, University of Tokyo, Kashiwa, Chiba 277-8581, Japan}

\author{Baile Zhang\orcidlink{0000-0003-1673-5901}}
\email{blzhang@ntu.edu.sg}
\affiliation{School of Physical and Mathematical Sciences, Nanyang Technological University, Singapore 637371, Singapore}
\affiliation{Centre for Disruptive Photonic Technologies, Nanyang Technological University, Singapore, 637371, Singapore}

\author{Y. D. Chong\orcidlink{0000-0002-8649-7884}}
\email{yidong@ntu.edu.sg}
\affiliation{School of Physical and Mathematical Sciences, Nanyang Technological University, Singapore 637371, Singapore}
\affiliation{Centre for Disruptive Photonic Technologies, Nanyang Technological University, Singapore, 637371, Singapore}

\begin{abstract}
  The surface states of certain topological phases can be linked to a quantum anomaly: the violation of a classical symmetry by a field theory via a non-conserved current.  This has been generalized to the case of a non-Hermitian (NH) chiral anomaly affecting the surfaces states of an NH Weyl phase.  Here, we show that the NH anomaly inflow is mediated by continnum Landau modes (CLMs): special eigenstates exhibiting both spatial localization and a continuous spectrum, contrary to the usual distinction between bound and free states.  The number of anomaly-induced surface modes scales with the sample volume rather than its surface area, which is shown to be tied to the unusual multiplicity of the CLMs.  The other properties of the CLMs, including their normalization conditions and localization scale, closely match the predictions of the NH field theory.   Finally, we discuss the conditions under which these phenomena can be probed experimentally using metamaterials.
\end{abstract}

\maketitle

\textit{Introduction}---For over four decades, topological bandstructures have been a topic of ongoing fascination, not only for their remarkable consequences (e.g., the quantization of the Hall conductance) but also their deep theoretical connections.  For example, it has been found that the bulk and edge states of topological materials can serve as manifestations of field theoretic anomalies (violations of classical symmetries by quantum fields \cite{fujikawa2004path, bertlmann2000anomalies}),  such as quantum Hall edge states and Weyl semimetal states with the chiral anomaly \cite{qi2008topological, jackiw1984fractional, haldane1988model, chiu2016classification, nenno2020axion, wang2024signature, fukushima2008chiral, zyuzin2012topological}, and fractional corner modes with the filling anomaly \cite{benalcazar2019quantization, xie2021higher, peterson2020fractional}.  Recently, it was discovered that the link between topological states and anomalies can be generalized to non-Hermitian (NH) topological phases \cite{bessho2021nielsen, kawabata2021topological}, which is remarkable since Hermiticity is usually a basic assumption for both band topology \cite{bergholtz2021exceptional, wang2023non, BanerjeeReview2023, Lee2016Anomalous, Leykam2017Edge, shen2018topological, gong2018topological} and field theory \cite{Bender2005, Bender2007Reports, Alexandre2017}.

NH topological phases are an important frontier for investigations of topological materials \cite{bergholtz2021exceptional, wang2023non, BanerjeeReview2023}, with growing interest in their realization using metamaterial platforms like photonic or acoustic lattices; indeed, non-Hermiticity is often an intrinsic and non-negligible feature of such platforms \cite{lin2011unidirectional, xue2022topological, weidemann2020topological, wang2022non, liu2024localization, helbig2020generalized, mcdonald2020exponentially, busnaina2024quantum, slim2024optomechanical}.  On the theoretical front, some NH bandstructures have been found to be characterizable by modifying existing Hermitian frameworks \cite{shen2018topological}, while others host intrinsically NH forms of band topology \cite{Lee2016Anomalous, Leykam2017Edge, shen2018topological, gong2018topological, kawabata2019symmetry, bergholtz2021exceptional, nakamura2024bulk}.  NH systems can also display novel and distinctive behaviors, one oft-cited example being the non-Hermitian skin effect (NHSE), a broad collapse of bulk states into boundary states that can be related to NH point gap topology \cite{hatano1996localization, wang2023non, zhang2022universal}.  

In this work, we show that the manifestation of an NH anomaly in a three-dimensional (3D) lattice occurs via unusual surface states called continuum Landau modes (CLMs) \cite{wang2023continuum, zhang2024observation}: eigenstates that form a continuous spectrum but are spatially localized, in violation of the usual dichotomy between bound and free states exhibited by Hermitian wavefunctions \cite{enss1983bound}.  CLMs were originally discovered in models with no evident connection to NH band topology \cite{wang2023continuum}, but here they emerge as surface states of a 3D NH Weyl semimetal (NHWSM) under an applied magnetic field.  We find that the number of surface states scales with the sample volume rather than its surface area---a phenomenon previously identified as an unusual form of NHSE \cite{bessho2021nielsen}, and now shown to follow from the combination of the NH anomaly \cite{kawabata2021topological} with the unusual multiplicity of the CLMs.  Consequently, the number of surface states can greatly exceed that of the Hermitian, area-scaling counterpart \cite{qi2008topological, chiu2016classification}.  Finally, we discuss how these exotic surface states may be experimentally probed using metamaterials.

\begin{figure}
    \centering
    \includegraphics[width=\linewidth]{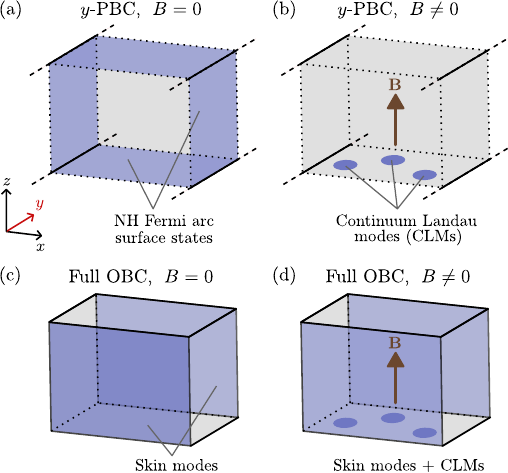}
    \caption{Surface states of a non-Hermitian Weyl semimetal with nonreciprocal couplings. (a) With periodic boundary conditions (PBC) along $y$, open boundary conditions (OBC) otherwise, and no external magnetic field, the $\pm x$ and $\pm z$ boundaries host surface states mapping to Hermitian Fermi arc surface states.  (b) A magnetic field applied along $+z$ turns the NH surface states into ``continuum Landau modes'' (CLMs) on the $-z$ boundary. (c) With OBC in all directions, all eigenstates collapse into skin modes on the $\pm y$ boundaries.  (d) When a magnetic field along $+z$ is further applied, the eigenstates are a mix of skin modes and CLMs.}
    \label{fig:schematic}
\end{figure}

\textit{Model}---We study a 3D tight-binding model comprising a Hermitian WSM with additional NH couplings \cite{kawabata2021topological, bessho2021nielsen, denner2021exceptional, nakamura2024bulk, zhang2022review}.  The NHWSM can be characterized by a 3D winding number \cite{bessho2021nielsen, sun2021geometric, kawabata2021topological} which is tied to a chiral magnetic effect  \cite{fukushima2008chiral, armitage2018weyl}: an applied \textit{magnetic} field can induce a particle flow in the direction of the magnetic field  \cite{kawabata2021topological, bessho2021nielsen}.  This is due to an NH anomaly amongst the two-dimensional surface states, whose complex dispersion relations $\pm k_x-ik_y$ break $U(1)$ axial symmetry.  By contrast, in the Hermitian $U(1)$ anomaly, an \textit{electric} field induces a polarization between left- and right-handed particles \cite{bertlmann2000anomalies}.

In a finite sample of the NHWSM, we find that the surface response is complicated by NH boundary effects, as summarized in Fig.~\ref{fig:schematic}.  Under periodic boundary conditions (PBC) along the $y$ direction and open boundary conditions (OBC) in other directions, and no applied magnetic field ($B = 0$), there are surface states with complex energies [Fig.~\ref{fig:schematic}(a)] that map to Fermi arc surface states of the Hermitian WSM \cite{kawabata2021topological, bessho2021nielsen}.  A magnetic field along the $+z$ direction turns these surface states into CLMs \cite{wang2023continuum} localized on the $-z$ surface [Fig.~\ref{fig:schematic}(b)].  On the $+z$ surface, the CLMs are non-normalizable, leading to an asymmetry in the density of surface states and hence the NH chiral magnetic effect \cite{kawabata2021topological}.  If the lattice is finite (OBC in all directions), the NHSE occurs \cite{bergholtz2021exceptional, hatano1996localization, wang2023non, zhang2022universal}.  For $B = 0$, all lattice modes collapse into skin modes on the $\pm y$ surfaces [Fig.~\ref{fig:schematic}(c)].  However, applying a magnetic field causes CLMs, distinct from skin modes, to re-emerge on the $-z$ surface [Fig.~\ref{fig:schematic}(d)].


\begin{figure*}
    \centering
    \includegraphics[width=1.0\linewidth]{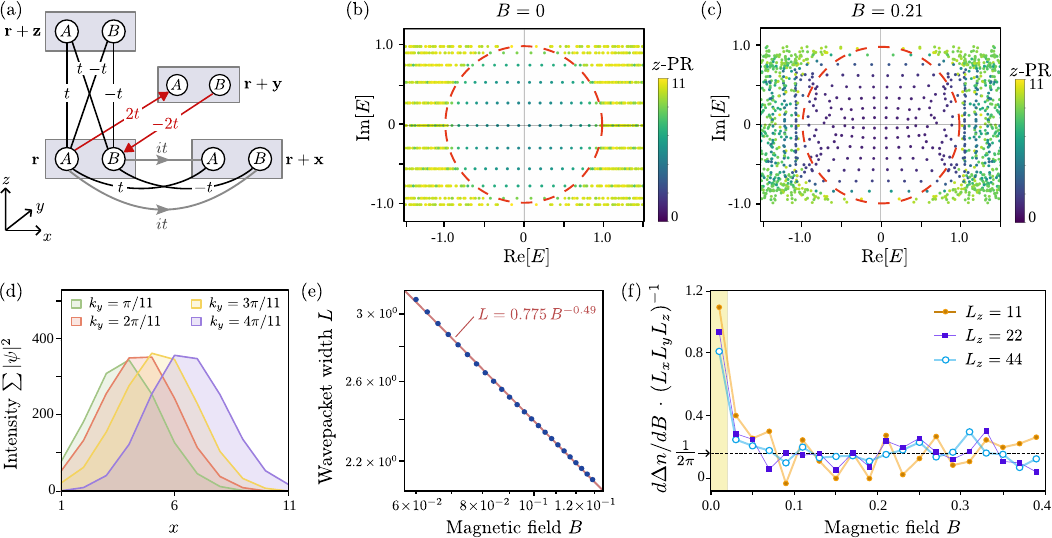}
    \caption{(a) 3D tight-binding model.  Each unit cell contains two sites $A$ and $B$, with mass $\pm \Delta$.  Adjacent cells are coupled by reciprocal hoppings $\pm t$ (black lines), nonreciprocal hoppings $\pm it$ (gray arrows; arrow direction indicates $+it$), and NH one-way hoppings (red arrows).  In subsequent plots, we set $t = 1/2$ and $\Delta = 2$.  (b) Complex spectrum for $y$-PBC and $B = 0$ [Fig.~\ref{fig:schematic}(a)], with lattice size $L_x = L_z = 11$, $L_y = 22$.  Colors denote the $z$ participation ratio (smaller means stronger localization).  The boundary of the point gap, $|E| = 1$, is marked by red dashes.  The states in $|E| < 1$ are NH Fermi arc surface states.  (c) Complex spectrum for $y$-PBC and $\mathbf{B} = 2\pi/30\, \hat{z}$ [Fig.~\ref{fig:schematic}(b)], with all other parameters the same as (b).  The states in $|E| < 1$ are now CLMs on the $-z$ surface.  (d) Surface state intensity $\sum |\psi|^2$ for a lattice of $L_x = L_z = 11$, with $y$-PBC.  The intensity is summed over sites along the bottom ($-z$) surface, both sublattices, and all eigenstates with $|E| < 1$; results are shown for various $k_y$, with each curve receiving contributions from multiple states of different $E$.  (e)  Log-log plot of the spatial width $L$ of a CLM (with $E \approx 0$) versus magnetic field $B$ (blue dots).  A least-squares fit of $\log L$ versus $\log B$ (red line) gives $L \propto 1/\sqrt{B}$ \cite{wang2023continuum}.  All other parameters are the same as in (b).
      (f) Plot of $d\Delta n/dB \cdot (L_xL_yL_z)^{-1}$ versus $B$. Here, $\Delta n = n_{-}- n_{+}$ is the difference in the number of in-gap modes ($|E|<1$) localized on the bottom and top surfaces, and the derivative is estimated using $\delta B = 0.02$.  Results are plotted using $L_x=L_y=11$, with different values of $L_z$.  The yellow-shaded region indicates the regime $L_x \lesssim 2B^{-1/2}$, where the system size is smaller than the CLM width.
    }
    \label{fig:clms}
\end{figure*}

\begin{figure*}
\centering
\includegraphics[width=1.0\textwidth]{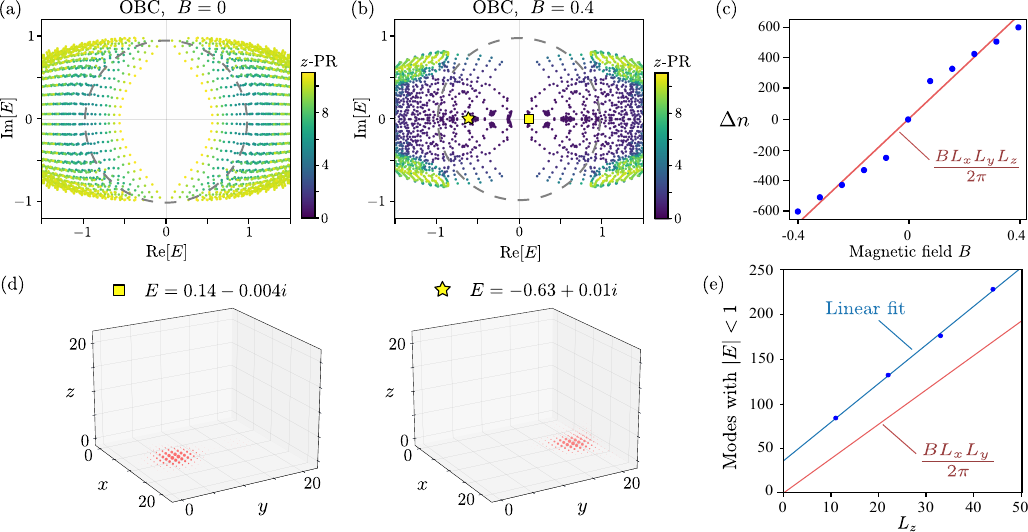}%
\caption{(a) Complex spectrum under full OBC, with colors indicating the $z$-PR of the modes.
(b) Complex spectrum under $B=0.4$. In-gap surface modes have a small $z$-PR as modes are pushed into the $-z$ plane. The point gap $|E|<1$ is indicated by the gray dashes.
(c) Mode imbalance $\Delta n$ versus $B$ (blue dots), calculated numerically using eigenstates with $|E|<1$. Solid line shows the estimate $\Delta n = B L_x L_y L_z /2\pi$.
  (d) Mode intensity profiles for the representative eigenstates labeled in (b), which are localized to points on the $-z$ surface.  The results for (a)--(d) are obtained with a $22\times22\times22$ lattice. (e) Number of modes with $|E|<1$ versus $L_z$, for $B=0.2$ and $L_x=L_y=11$ (blue dots).  The blue line is the linear least-squares fit. 
}
\label{fig:full OBC}
\end{figure*}
 
The model is depicted in Fig.~\ref{fig:clms}(a), and has the tight-binding
equations
\begin{align}
  -\Delta\, \psi^A_{\mathbf{r}}
  + \sum_{\pm} \left[\,t \left(\psi^A_{\mathbf{r}\pm \mathbf{x}}
    + \psi^A_{\mathbf{r}\pm \mathbf{z}} \mp \psi^B_{\mathbf{r} \pm \mathbf{z}}\right)
    \mp it \psi^B_{\mathbf{r}\pm \mathbf{x}} \right]\;\;\;\nonumber\\
  + 2t \psi^A_{\mathbf{r}-\mathbf{y}} = E \psi^A_{\mathbf{r}}, \label{NH_WSM1} \\
  \Delta \psi^B_{\mathbf{r}}
  + \sum_{\pm} \left[ -t \left(\psi^B_{\mathbf{r}\pm \mathbf{x}}
    + \psi^B_{\mathbf{r}\pm \mathbf{z}\mathbf{}} \mp \psi^A_{\mathbf{r} \pm \mathbf{z}}\right)
    \mp it \psi^A_{\mathbf{r}\pm \mathbf{x}} \right]\;\;\;\nonumber \\
  - 2t \psi^B_{\mathbf{r}+\mathbf{y}} = E \psi^B_{\mathbf{r}},
  \label{NH_WSM2}
\end{align}
where $A$ and $B$ denote sublattices of a cubic lattice (with unit lattice vectors $\{\mathbf{x},\mathbf{y},\mathbf{z}\}$), $\psi^{A/B}_{\mathbf{r}}$ is the wavefunction on site $A$/$B$ at cell position $\mathbf{r}$, $E$ is the eigenenergy, $t$ is a hopping parameter, and $\Delta$ is a sublattice-specific detuning.  There are Hermitian reciprocal couplings along $x$ and $z$ (black lines), Hermitian nonreciprocal couplings along $x$ (gray arrows), and one-way couplings along $y$ (red arrows).  The one-way hoppings are the only NH parts of the model, but the Hermitian part still breaks time-reversal symmetry due to the $\pm it$ hoppings \cite{kawabata2021topological, bessho2021nielsen}.

Under $y$-PBC, the Hamiltonian has Hermitian and anti-Hermitian parts $H_R + i H_I$, where $H_I = -2t\sin{k_y}$, regardless of the boundary conditions in the other directions.  If $\Delta \in [2t,\space 6t]$, which we shall assume in all of the following, $H_R$ is a T-broken WSM \cite{armitage2018weyl}: when we take PBC in all directions, there are two opposite-chirality Weyl points at $\mathbf{k}_\pm = (0, \pm k_0, 0)$ where $k_0 = \cos^{-1}[ {\Delta/2t -2}]$.  The full NH Hamiltonian has the same bulk eigenstates; expanding it around the original Weyl points, with $\mathbf{k} = \mathbf{k}_\pm + \mathbf{q}$, yields
\begin{multline}
H_{\pm}(\mathbf{q}) = 2t \big[\mp i \sin(k_0) - i  \cos(k_0) q_y + \sigma_x q_x  \\
\quad \mp  \sin (k_0) \sigma_z q_y + \sigma_y q_z \big].
\label{NH_Weyl_dispersion}
\end{multline}
These have the same eigenstates as the WSM, but different bulk spectra \cite{kawabata2021topological, sun2021geometric, denner2021exceptional, bessho2021nielsen, rui2022non} that are complex-valued and have point gaps \cite{kawabata2019symmetry, bergholtz2021exceptional, hatano1996localization, wang2023non}.  As shown in the Supplemental Materials \cite{Supplementary}, for $\Delta=4t$ the point gap's boundary is the circle $|E| = 2t$.


If we truncate the lattice with OBCs, retaining the $y$-PBC, the Hermitian WSM has Fermi arc surface states, as depicted in Fig.~\ref{fig:schematic}(a).  These are also eigenstates of the NH model, but with complex $E$ in the point gap, as verified numerically in Fig.~\ref{fig:clms}(b).  Hereafter, we take $t=1/2$ and $\Delta = 2$, so that $\mathbf{k}_\pm = (0, \pm \pi/2, 0)$ and the point gap is $|E| < 1$.  Each data point in Fig.~\ref{fig:clms}(b) is colored by the participation ratio along the $z$ direction, defined as
$z\textrm{-PR} \equiv \left(\sum_{z} I_z\right)^2/\sum_{z} I_z^2$ where $I_z = \sum_{x,y} |\psi_{\mathbf{r}=(x,y,z)}|^2$ \cite{Thouless1974}; for $|E| < 1$, the $z$-PR is indeed much lower (i.e., these states are localized along the $z$ direction).

We now apply a uniform magnetic field via a vector potential $\mathbf{A} = Bx \mathbf{y}$ (with $B > 0$), which enters the phases of the one-way $y$-hoppings.  In a Hermitian WSM, a magnetic field creates Weyl orbits by hybridizing Fermi arc states with bulk Landau states \cite{potter2014quantum, zhang2016quantumoscillation}.  The NH behavior is completely different, since $[H_R, H_I] \ne 0$ when $B \ne 0$.  Under $y$-PBC, the NH lattice exhibits localized surface states on the $-z$ boundary, but no surface states on the other boundaries [Fig.~\ref{fig:schematic}(b)]. Figure~\ref{fig:clms}(c) plots the complex energies and $z$-PR for $B = 0.21$.  The localized states' energies still mostly lie in $|E| < 1$, but their wavefunctions are now localized around a point on the surface, not extended over the surface.  They can be identified as CLMs, which were first studied in a 2D NH model with dispersion relation $E = \pm k_x - ik_y$ \cite{wang2023continuum}---exactly matching the $B = 0$ surfaces states of the NHWSM.

The CLMs are exotic NH states that violate the usual dichotomy between discrete bound states and continuum free states.  In a 2D infinite continuous space under $y$-PBC, CLMs form a continuous family of Gaussian wavepackets of width $\sim B^{-1/2}$ and center $x_0=[\text{Im}(E)+k_y]/B$, where $E$ spans $\mathbb{C}$ \cite{wang2023continuum}.  In the NHWSM, the surface states are limited to the point gap, $|E| < 1$, and for $B > 0$ exist only on the $-z$ surface (being non-normalizable on the $+z$ surface).  Figure~\ref{fig:clms}(d) shows the $x$-dependence of the surface intensity $\sum |\psi|^2$, summed over the $-z$ surface sites for all states in $|E| < 1$.  For each $k_y$, we find Gaussian-like profiles with $x$-centers varying linearly with $k_y/B$.  Importantly, each profile comprises \textit{many} CLMs, centered around the same $x$ with spatial width $\sim B^{-1/2}$ [Fig.~\ref{fig:clms}(e)], but different $E$.  This is tied to the fact that CLMs are not mutually orthogonal  \cite{bergholtz2021exceptional, wang2023non, BanerjeeReview2023}.  By contrast, Hermitian Landau levels satisfy $x_0 \propto k_y/B$ with \textit{one} orthogonal mode per $k_y$.

This has important consequences for the NH chiral magnetic effect \cite{fukushima2008chiral, bessho2021nielsen, kawabata2021topological}.  Notably, the anomaly-induced surface-mode imbalance scales with the sample \textit{volume}, rather than the surface area.  This is shown in Fig.~\ref{fig:clms}(f), where we calculate $\Delta n \approx n_- - n_+$ (where $n_{\pm} = \sum |\psi_{\pm}|^2$, with the sum taken over all $k_y$ under $y$-PBC, all modes in the point gap, and sites on the $\pm z$ surface), and plot $d\Delta n / dB \cdot (L_xL_yL_z)^{-1}$ versus $B$ for $L_z \in \{11, 22, 44\}$.  This volume scaling can also be inferred from the bulk topological invariant, which yields \cite{bessho2021nielsen}
\begin{equation}
  \Delta n \sim \frac{B\, L_x L_y L_z}{2\pi}.
  \label{anomaly}
\end{equation}
Due to the $L_z$-dependence, this was identified in Ref.~\onlinecite{bessho2021nielsen} as an unusual form of NHSE directed along $z$ (distinct from the standard NHSE, which also occurs in this model; see below). We now see that Eq.~\eqref{anomaly} arises microscopically from the high multiplicity of CLMs at each $x_0$, together with the Landau-level-like guiding-center spacing $\delta x_0 = \delta k_y/B$ (which produces the factor $BL_xL_y/2\pi$).
Note also that in Fig.~\ref{fig:clms}(f), the volume-scaling holds for $L_x \gtrsim 2B^{-1/2}$---i.e., sample width large relative to the CLM widths induced by $B$.


\begin{figure}
    \centering
    \includegraphics[width=1.0
    \linewidth]{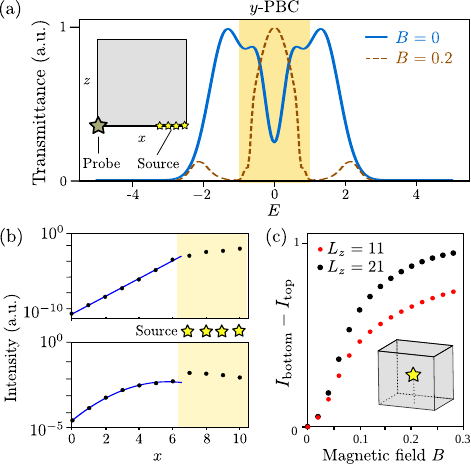}
    \caption{(a) Transmittance spectrum on the $-z$ surface of a $y$-PBC sample, with source and probe on the sides of the $-z$ surface (inset schematic).  For $B=0$ (blue line), transmission is nonzero over a range surrounding the point gap $|E|<1$ (yellow region). For $B=0.2$ (brown dashes), a strong peak develops in the point gap, corresponding to CLM surface states.  Each curve is normalized to a maximum transmittance of 1.  (b) Field intensity $I$ on the $-z$ surface for $B=0$ and $B=0.2$. The last four sites are excited by point sources.  For $B=0$, the field decays exponentially away from the excitation (top plot), but for $B=0.2$ the decay is Gaussian.  Solid lines show linear and quadratic fits of $\log I$ versus $x$, excluding the excitation points.  These calculations use $y$-PBC with $k_y=0$ and $L_x=L_z=11$.  (c) Intensity difference between bottom and top halves of the lattice, versus magnetic field $B$, for an OBC lattice with $L_x=L_y=21$ and $L_z = 11, \space 21$.  The source is placed in the center of the lattice (see inset). }
    \label{fig:phenomena}
\end{figure}

Many of these features carry over to the case of a finite sample, with OBC in all directions.  For $B = 0$, all lattice eigenstates are skin modes localized along $\pm y$, due to the NHSE \cite{bergholtz2021exceptional, wang2023non, zhang2022universal}.  The complex energy spectrum, plotted in Fig.~\ref{fig:full OBC}(a), differs qualitatively from the $y$-PBC spectrum of Fig.~\ref{fig:clms}(b).  The point gap region $|E| < 1$ is no longer densely filled, and the eigenstates that do occur in this region are localized along the $\pm y$ boundaries like all the other skin modes.  The NHSE apparently destroys the NH Fermi arc states \cite{zhang2022universal}.

Remarkably, however, the surface states re-emerge when a magnetic field is applied.  The complex spectrum for $B=0.4$ is plotted in Fig.~\ref{fig:full OBC}(b), revealing that many states with low $z$-PR have returned into the $|E| < 1$ region.  In Fig.~\ref{fig:full OBC}(c), we plot the mode imbalance $\Delta n$ for $|E| < 1$ states, and find that the variation with $B$ is linear and closely follows Eq.~\eqref{anomaly}, indicating the re-emergence of NH anomaly inflow.  These eigenstates are localized to different points on the $-z$ surface, as shown by the two representative intensity profiles plotted in Fig.~\ref{fig:full OBC}(d).  (Some states with $|E| > 1$ are also seen to have small $z$-PR in Fig.~\ref{fig:full OBC}(b), but they turn out to be hinge modes, not surface modes.)  As in the $y$-PBC case, the number of surface states scales linearly with $L_z$ [Fig.~\ref{fig:full OBC}(e)], and their widths scale as $B^{-1/2}$ along $x$ and $y$ \cite{Supplementary}.

Finally, we suggest how these features can be probed experimentally.  NH lattices have been implemented and studied on many classical-wave platforms \cite{lin2011unidirectional, xue2022topological, weidemann2020topological, wang2022non, liu2024localization, helbig2020generalized, mcdonald2020exponentially, busnaina2024quantum, slim2024optomechanical}, some of which allow for both non-Hermiticity and T-breaking \cite{liu2024localization}.  Putting aside implementation-specific details, we use the tight-binding model as a guide to demonstrate how the underlying theoretical phenomena may manifest in transmission measurements.  For these calculations, we apply an additional uniform loss term of $-2it$ to every site to prevent signal blowup.  When a weakly-connected source with energy (frequency) $E$ and spatial profile $\left| \psi_{\mathrm{in}} \right\rangle$ is connected to the lattice, it produces a steady-state field
\begin{equation}
  \left| \psi_{\textrm{out}} \right\rangle = 
  \left(E - \hat H'\right)^{-1}
  \left| \psi_{\textrm{in}} \right\rangle,
  \label{response}
\end{equation}
where $\hat{H}'$ is the lattice Hamiltonian including the uniform loss.
At any given site, the transmittance measured by a weakly-connected probe is $T(E) = |\psi_{\textrm{out}}|^2$, the un-normalized absolute square of the wavefunction.  Figure~\ref{fig:phenomena}(a) shows the transmission spectrum for a $y$-PBC sample (with $k_y=0$), with source and probe sites on the $-z$ plane.  Here, we use a 4-site excitation source to ensure more robust coupling to extended lattice modes \cite{wetter2022observation, cantillano2017observation}.  For $B = 0$, the transmittance curve (blue line) forms peaks around $|E|=1$ and is elevated over much of the point gap region $|E| \lesssim 1$, consistent with the excitation of NH Fermi arc surface states.  For $B=0.2$, the transmittance peaks much more strongly inside the point gap (brown dashes).  Note that these two transmittance curves are individually normalized to peak at 1 in order to show the $E$-dependence more clearly; using the same normalization for both curves, the $B = 0.2$ transmittance is substantially higher (see below). 

The emergence of CLMs for $B \ne 0$ can also be pinpointed through their characteristic Gaussian spatial profiles.  In Fig.~\ref{fig:phenomena}(b), we plot the transmitted field on the $-z$ surface, using the same 4-site excitation as before (with frequency $E=0$). When $B=0$, the field decays exponentially, as shown in the upper plot, as expected of ballistic states (NH Fermi arc states) with added loss. However, for $B=0.2$ the transmission has a Gaussian variation, matching the profile of the CLMs.

Moreover, we can obtain evidence for the NH chiral magnetic effect, including the $L_z$-dependence found in Eq.~\eqref{anomaly}, by studying how the transmittance varies with $B$ and $L_z$.  We consider placing an excitation source in the center of a sample with full OBC, operating at $E = 0$, and find the transmittances to the lower and upper halves of the sample (i.e., summing over all sites in each half).  In Fig.~\ref{fig:phenomena}(c), we plot the difference in total transmittance between the lower and upper sides, which is found to increase with both $B$ and $L_z$.

\textit{Conclusions}---The surface of an NHWSM hosts rich behaviors tying together the recently-discovered phenomena of CLMs, non-Hermitian anomalies, and the NHSE.  Under $y$-PBC, the Fermi-arc surface states are converted by a magnetic field $B$ into CLMs, generating a particle imbalance associated with the NH anomaly. The resulting volume scaling is shown to originate from the unusual multiplicity of the CLMs.  Under full OBC, the skin effect collapses the bulk and surface modes, but a magnetic field causes the CLMs to re-emerge.  These behaviors should be observable on existing experimental platforms such as fiber loop-based synthetic lattices \cite{weidemann2020topological}, gyromagnetic metamaterials \cite{liu2022topological}, acoustic resonators \cite{gao2022anomalous}, or electric circuits \cite{helbig2020generalized}.

This work was supported by the Singapore National Research Foundation (NRF) under the NRF Investigatorship NRF-NRFI08-2022-0001, and Competitive Research Program (CRP) Nos.~NRF-CRP23-2019-0005, NRF-CRP23-2019-0007, and NRF-CRP29-2022-0003.  K.~K.~is supported by MEXT KAKENHI Grant-in-Aid for Transformative Research Areas A ``Extreme Universe" No.~JP24H00945.

\bibliography{3DNH_references}

\clearpage

\begin{widetext}

\appendix

\onecolumngrid

\renewcommand{\theequation}{S\arabic{equation}}
\setcounter{equation}{0}

\counterwithout{figure}{section}
\newcounter{suppfigure}
\counterwithin{figure}{suppfigure}
\renewcommand{\thefigure}{S\arabic{suppfigure}}

\begin{center}
  {\large \textbf{Supplemental Materials for}}\\
  \vskip 0.05in
  {\Large ``Continuum Landau surface states in a non-Hermitian Weyl semimetal''}\\
  \vskip 0.1in

  Shuxin Lin, Rimi Banerjee, Zheyu Cheng, Kohei Kawabata, Baile Zhang, and Y. D. Chong
\end{center}

\section*{\textbf{Section A: 2D NH Dirac Dispersion and CLMs}}

Here, we briefly review the continuum CLM model \cite{wang2023continuum}.  For a 2D NH Dirac dispersion, consider the Hamiltonian
\begin{equation}
    H = \mp i \frac{\partial}{\partial x} + \frac{\partial}{\partial y} , \label{H_2DNH}
\end{equation}
whose eigenvectors are plane waves of the form 
\begin{equation}
    \psi = C\exp{(ik_x x - ik_y y)}, \label{2}
\end{equation}
corresponding to a continuum spectrum
\begin{equation}
    E_0(k_x, k_y) = \pm k_x -ik_y, \label{3}
\end{equation}
These eigenvectors are free (extended) states with a continuous spectrum.  When a uniform magnetic field is applied, the story changes.  The Hamiltonian becomes
\begin{equation}
  H_{\textrm{CLM}} = s\left( -i \frac{\partial}{\partial x}  \right)   -i\left( -i \frac{\partial}{\partial y} - Bx \right), \label{H_CLM}
\end{equation}
where $s = \pm 1$.  Here we have introduced the Landau gauge $\mathbf{A} = (0,\space  Bx,\space 0)$, which corresponds to a magnetic field $\mathbf{B} = B \mathbf{z}$, with $e=c=1$.  Note that we have adopted a different sign convention from Ref.~\onlinecite{wang2023continuum}, equivalent to flipping the sign of the electric charge $e$.  The Hamiltonian is translationally invariant along $y$, so we look for eigenfunctions of $H_{\textrm{CLM}}$ of the form $\varphi(x, k_y) \, \exp(ik_y y)$.  We then obtain
\begin{equation}
\begin{aligned}
   \varphi(x, k_y)   = C \exp \left[ -\frac{B}{2} \left(x - \frac{\text{Im}(E)+k_y}{B}\right)^2   - i \text{Re}(E) \cdot x - ik_y \cdot y \right],
\end{aligned}
\label{psi_CLM}
\end{equation}
which describes a Gaussian wavepacket centered at
\begin{equation}
  \begin{aligned}
    x_0 = \frac{\text{Im}(E)+k_y}{B},
  \end{aligned}
  \label{energy-position}
\end{equation}
and with
\begin{equation}
  k_x = -\mathrm{Re}(E). \label{CLM_kx}
\end{equation}

This energy-position relation is reminiscent of Hermitian Landau levels.  Recall that for each Landau level, the guiding center is $x_{0} = k_y/B$.  This one-to-one correspondence between $k_y$ and the $x$-center is helpful for enumerating the density of states.  In a sample of size $L_x\times L_y$, the center of the ``cyclotron orbit'' must remain within the sample (e.g., $0 \le x_0 \le L_x$), while the momentum along $y$ is discretized to multiples of $2\pi / L_y$.  Hence, the number of states is
\begin{equation}
  n_{\textrm{LL}} = \frac{B L_x L_y}{2\pi}.
  \label{LL_scaling}
\end{equation}
However, there is a critical difference between Landau levels and CLMs.  In the CLM solution, Eq.~\eqref{psi_CLM}, $k_x$ is an additional independent quantum number---indexing different CLMs, with different $\mathrm{Re}(E)$---for each choice of $k_y$ (and $x_0$).  As we show in the main text, this produces extra multiplicity that leads to CLM surface states deviating from the area-scaling rule of Eq.~\eqref{LL_scaling}.

\section*{\textbf{Section B: Point gap as criteria for surface modes}}

\setcounter{suppfigure}{3}

The non-Hermitian Weyl semimetal (NHWSM) model we study is described by the bulk momentum-space Hamiltonian
\begin{equation}
    H = 2t  \sigma_x \sin k_x -2it\sin k_y + 2t  \sigma_y \sin k_z  + 
 \sigma_z( -\Delta + 2t \cos k_x + 2t \cos k_y + 2t \cos k_z ).  \label{k-space_NH_WSM}
\end{equation}
Its energy spectrum is
\begin{align}
  E_{\pm} &= -2it \sin k_y \pm |\mathbf{a}|, \\
  \mathbf{a} &=
  2t\left(\sin k_x, \space \sin k_z, \space -\Delta/2t + \cos k_x + \cos k_y + \cos k_z\right).
\end{align}
This exhibits a point gap for $2t<\Delta<6t$.  To prove this, let us take $t = 1/2$ and define
\begin{equation}
  f = |E|^2
  = 3 + \Delta^2 - 2\Delta\left(\cos k_x + \cos k_y + \cos k_z\right)
  + 2\cos k_x \cos k_y + 2\cos k_y \cos k_z + 2\cos k_z \cos k_x.
\label{Spectra}
\end{equation}
We now perform the second derivative test on $f(\mathbf{k}) $ to find local maxima and minima. For $\Delta \rightarrow 2^{+}$, at $\mathbf{k} = (0,0,\pi)$ we have $f = (\Delta-1)^2 \rightarrow 0^{+}$. For $\Delta \rightarrow 6^{-}$, at $\mathbf{k} = (0,0,0)$ we have $f = (\Delta-3)^2 \rightarrow 0^{+}$.

In the case $\Delta = 2$, corresponding to the parameters we use in the main text, $\nabla f = 0$ is satisfied at points along the lines $\mathbf{k} = (k_x,0,0)$, $\mathbf{k} = (0,k_y,0)$, and $\mathbf{k} = (0,0,k_z)$, corresponding to a value of $f = 1$. We can check that $|E|=1$ is the global minimum, either numerically or by comparing analytically to the values on the boundary of the Brillouin zone.  Note that the circularity of the point gap holds only for this particular choice of parameters; for other parameters, the point gap still exists (provided $2t<\Delta<6t$) but may be non-circular.

The NHWSM can be shown to exhibit a 3D anomaly in the bulk, which is compensated by a 2D anomaly on the surfaces \cite{kawabata2021topological}. 
It is defined by a nontrivial winding number when $\text{det}(H-E) \neq 0$:
\begin{equation}
  \begin{split}
    W_3(E) = -\int_{BZ} \frac{d^3k}{24\pi^3} \epsilon^{i j k} \mathrm{tr} \left[ (\tilde{H}^{-1} \partial_{k_i} \tilde{H}) \right. 
\left. \times (\tilde{H}^{-1} \partial_{k_j} \tilde{H}) (\tilde{H}^{-1} \partial_{k_k} \tilde{H}) \right].
  \end{split} 
\end{equation}
This 3D winding number, which vanishes for Hermitian systems \cite{bessho2021nielsen}, counts the chirality of long-time Weyl points inside the NHWSM.  The Weyl points at $k_y = \pm \pi/2$ with chirality $\pm 1$ have complex energies $\mp i$, and only the one with chirality $-1$ survives at long times. Thus, adding a magnetic field $\mathbf{B} = (0 ,0 ,B)$ is predicted to induce a Landau channel with $z$-directional velocity antiparallel to $\mathbf{B}$, forming a bulk current antiparallel to $\mathbf{B}$. For a finite sample, the top surface loses modes and the bottom surface gains modes, compensating the bulk current \cite{kawabata2021topological}. To compensate this bulk current, NH field theory \cite{kawabata2021topological} predicts a number of surface states $\textit{per energy}$ $E$:

\begin{equation}
  \begin{split}
    n_{bot}(E) = +\frac{\Phi}{2\pi}, \quad n_{top}(E) = -\frac{\Phi}{2\pi} .
  \end{split} 
\end{equation}
Where $\Phi$ is the magnetic flux going through the area cross section. Intuitively, this is the result of the NH chiral magnetic effect induced by the 3D winding number, that allows a 1D skin effect along the direction of the magnetic field. As a result, the $\textit{total}$ number of surface states should scale with sample thickness $L_z$. The scaling of $L_z$ does not enter through the spectral density $n(E)$, but instead increases the number of viable energies $E$.

\section*{\textbf{Section C: Numerical studies of surface CLMs}}

\setcounter{suppfigure}{1}
\begin{figure}
    \centering
    \includegraphics[width=1.0\linewidth]{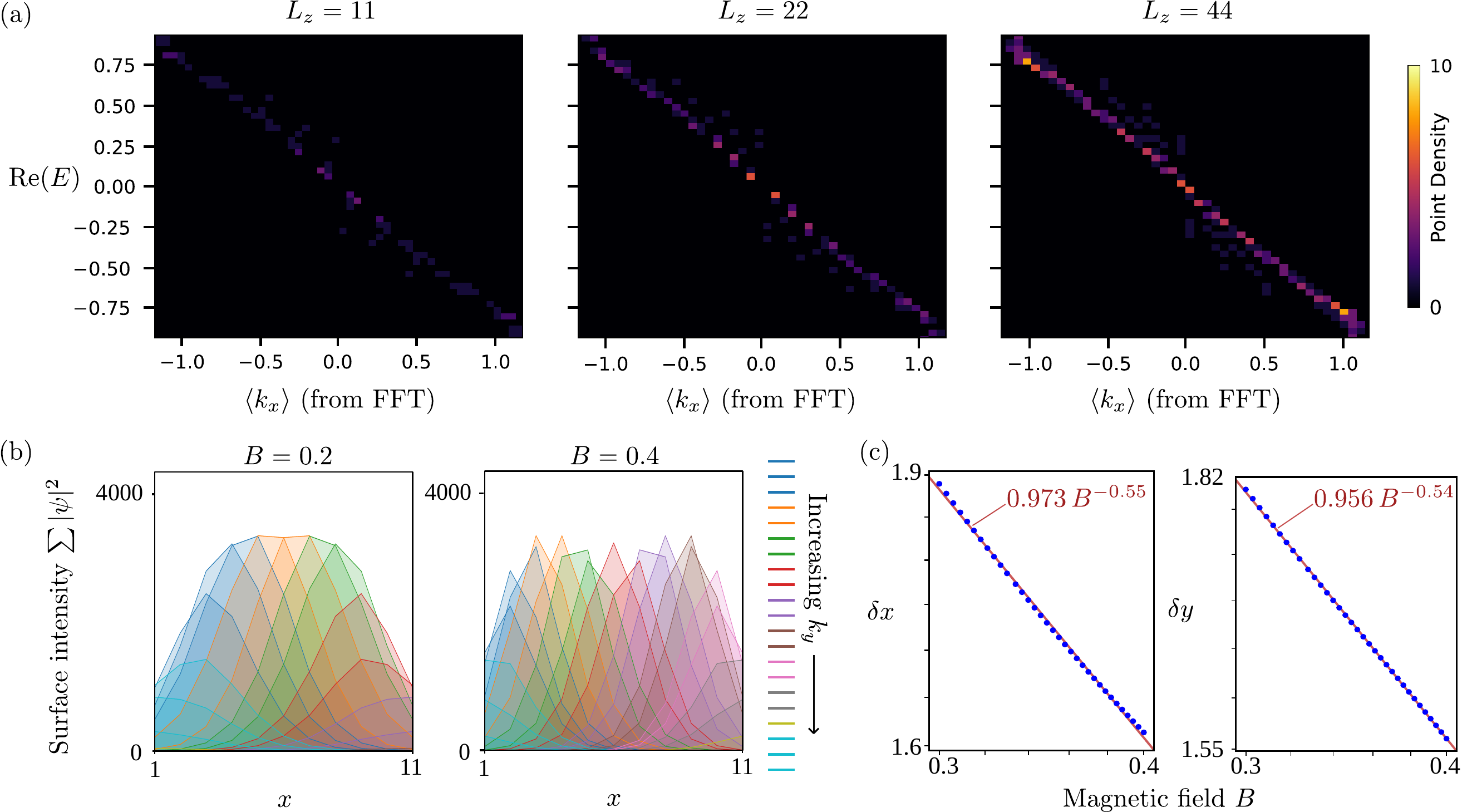}
    \caption{(a) Numerical results illustrating the energy-position relation for CLM surface states of the NHWSM.  For $y$-PBC lattices of size $L_x=L_y=11$, with different $L_z$, and applied magnetic field $B = 0.21$, we extract the eigenstates with $|E| < 1$, find $\langle k_x\rangle$ by Fourier transforming in $x$, and plot $\text{Re}E$ versus $\langle k_x\rangle$. (b) Distribution of CLM surface states, resolved by $k_y$.  For $L_x = 11$ and $L_z = 22$, with two choices of $B$, we show the $x$-dependence of the surface intensity $\sum |\psi|^2$, where the sum is taken over sites on the bottom ($-z$) surface for all $|E| < 1$ eigenstates.  With increasing $B$, more Gaussian wavepackets (corresponding to different $k_y$ fit in the sample, giving rise to an increase in the density of surface states; this is similar to Landau levels, but with a higher multiplicity of states for each $k_y$. (c) Variation of the wavepacket width versus $B$ in a full-OBC sample.  The widths along the $x$ and $y$ directions, $\delta x$ and $\delta y$, are extracted from a representative surface state (initially identified at energy $E_0 = 0.155 - 0.063i$ for $B=0.4$) and tracked as $B$ varies.  The system size is $L_x= L_y = L_z=11$, with OBC applied in all directions.  The semilogarithmic best-fit curves, shown in red, agree with the predicted $B^{-1/2}$ scaling.}
    \label{supp_fig:counting}
\end{figure}

In this section, we provide more details about how we count the surface states in the Non-Hermitian Weyl semimetal (NHWSM) lattice model.

First, we study NHWSM lattices with OBC in the $x$ and $z$ direction, and PBC in the $y$ direction.  We use different choices of sample thickness $L_z$, with fixed $L_x=L_y=11$ and $B =  0.21$.  After calculating the eigenstates of each lattice numerically, we focus on those with energies inside the point gap $|E|<1$; as explained in the main text, these states are verified to be surface states on the $- z$ surface (since $B > 0$).  

We now Fourier transform each surface state in the $x$ direction, and obtain the mean momentum $\langle k_x\rangle$.  Hence, we can plot the distributions of $\text{Re}E$ and $\langle k_x\rangle$, which are shown in Fig.~\ref{supp_fig:counting}(a).  The results show that $\text{Re}(E) \propto -\langle k_x\rangle$, consistent with the energy-position relation \eqref{CLM_kx} predicted for CLMs.  Notice also that increasing $L_z$ causes the distribution to become denser, indicating that more surface states are participating.  This is the source of the volume-scaling relation discussed in the main text.  We can reinterpret this scaling rule as applying to \textit{each} complex energy $E$ (in the limit of a semi-infinite sample); in the NH lattice, $L_z$ states contribute, giving rise to volume-scaling.

To gain additional insight into how additional CLM surface states emerge with increasing $B$, we resolve the eigenstates by $k_y$.  In Fig.~\ref{supp_fig:counting}(b), we show the $x$-dependence of the surface intensity (obtained by summing $|\psi|^2$ over sites on the $-z$ surface, for all states with $|E| < 1$, similar to Fig.~2(d) of the main text).  It can be seen that the intensities form Gaussian-like profiles, sorted by $k_y$.  With increasing $B$, the $x$-width decreases, and profiles with a larger range of $k_y$ fit into the sample, similar to the Landau level mechanism giving rise to the scaling relation in Eq.~\eqref{LL_scaling}.  However, we emphasize that each colored curve in Fig.~\ref{supp_fig:counting}(b) consists of \textit{many} CLMs, whereas for Landau levels there is one state per $k_y$.

As discussed in the main text, the surface CLMs also emerge when a magnetic field is applied to a sample with full OBC; this is somewhat remarkable because, in the $B = 0$ limit, the non-Hermitian skin effect (NHSE) causes all modes to collapse into skin modes on the $\pm y$ surfaces.  We have verified numerically, in different ways, that the surface states on the $-z$ surface under full OBC are indeed CLM-like.  In the main text, we show that the mode imbalance follows $\Delta n \propto BL_z$ scaling [Fig.~3(c) and (e)], and that individual modes are localized around points on the $-z$ surface [Fig.~3(d)].  As a further check, we take a state localized at the $-z$ surface (with energy nearest to $E=0$) for $B = 0.4$, and track this state as $B$ is gradually decreased. As shown in Fig.~\ref{supp_fig:counting}(c), the variation of the spatial width is close to the predicted $B^{-1/2}$ scaling for CLMs.

\clearpage
\end{widetext}
\end{document}